\documentclass[aps,prb,twocolumn,superscriptaddress]{revtex4}
\usepackage{graphicx}
\usepackage{amssymb}
\usepackage{amsmath}
\usepackage{color}
\usepackage{ulem}

\usepackage{pdfpages}
\begin{document}

\title{Schwinger boson approach for the dynamical mean-field theory of the Kondo lattice}

\author{Rulei Han}
\affiliation{Beijing National Laboratory for Condensed Matter Physics and
Institute of Physics, Chinese Academy of Sciences, Beijing 100190, China}
\affiliation{University of Chinese Academy of Sciences, Beijing 100049, China}
\author{Danqing Hu}
\affiliation{Beijing National Laboratory for Condensed Matter Physics and
Institute of Physics, Chinese Academy of Sciences, Beijing 100190, China}
\author{Jiangfan Wang}
\affiliation{Beijing National Laboratory for Condensed Matter Physics and
Institute of Physics, Chinese Academy of Sciences, Beijing 100190, China}
\author{Yi-feng Yang}
\email[]{yifeng@iphy.ac.cn}
\affiliation{Beijing National Laboratory for Condensed Matter Physics and
Institute of Physics, Chinese Academy of Sciences, Beijing 100190, China}
\affiliation{University of Chinese Academy of Sciences, Beijing 100049, China}
\affiliation{Songshan Lake Materials Laboratory, Dongguan, Guangdong 523808, China}

\date{\today}

\begin{abstract}
We apply the dynamical large-$N$ Schwinger boson technique as an impurity solver for the dynamical mean-field theory (DMFT) calculations of the Kondo lattice model. Our approach captures the hybridization physics through the DMFT self-consistency that is missing in the pure Schwinger boson calculations with independent electron baths. The resulting thermodynamic and transport properties are in qualitative agreement with more rigorous calculations and give the correct crossover behavior over a wide temperature range from the local moment regime to the Fermi liquid. Our method may be further extended to combine with the density functional theory for efficient material calculations.
\end{abstract}

\maketitle

\section{Intorduction}
Heavy fermion systems are typically described by the Kondo lattice model (KLM), which consists of an array of localized $f$-moments coupled with background conduction electrons \cite{Hewson1997,Coleman2005_a}. The localized-to-itinerant transition of the $f$-electrons underlies many anomalous low-temperature behaviors. Various approaches have been exploited to investigate the mechanism of the transition. Among them, the dynamical mean-field theory (DMFT) \cite{Metzner1989,Georges1992,Georges1996}, which maps the lattice model onto an effective impurity model, provides a simple yet powerful way to study the Kondo lattice physics \cite{Costi2002,Otsuki2009_a,Otsuki2009_b,Bodensiek2011,Matsumoto1995}. In practice, the DMFT impurity solver plays a major role for its implementation. Each impurity solver has its own pros and cons. Continuous-time quantum Monte Carlo (CT-QMC) \cite{Gull2012} requires analytic continuation to obtain the real-frequency information; the numerical renormalization group (NRG) \cite{Wilson1975,Bulla2008} gives the real-frequency spectral function but is computationally more expensive and difficult to apply for multiple orbitals. Compared to these exact numerical approaches, the slave boson representation has high computational efficiency but its application with the non-crossing approximation (NCA) may give unphysical results at low temperatures \cite{Kim1990,Palsson2001}.

Recently, a Schwinger boson representation \cite{Arovas1988} for the Kondo model has been developed to describe the local moment magnetism, but it has not been implemented within the DMFT framework. Parcollet and Georges first applied it to analyze the multichannel Kondo impurity model and found that it preserves the Fermi-liquid nature of the model in the exactly screened case \cite{Parcollet1997}. Later, Rech \emph{
et al.} applied it to the two-impurity model and obtained the ``Varma-Jones" fixed point \cite{Rech2006}. It has also been extended to the KLM \cite{Komijani2018,Komijani2019,Wang2020_a,Wang2020_b} and the Hund's metal \cite{Drouin-Touchette2021} and used to describe the strange metal in CeRh$_6$Ge$_4$ \cite{Shen2020} and the metallic spin liquid phase in CePdAl \cite{Zhao2019}. These successes motivate us to explore the possibility of its implementation as an impurity solver of DMFT for future combination with realistic material calculations \cite{Kotliar2006}.

In this work, we make the first attempt to combine the Schwinger boson impurity solver into the DMFT framework to study the paramagnetic normal phase of the KLM. Compared with the Abrikosov fermion representation \cite{Read1983,Auerbach1986}, which yields an artificial phase transition on the mean-field level, the Schwinger boson approach describes the Kondo physics through an auxiliary fermionic holon field and captures correctly the crossover between the local moment and Fermi liquid states. Within the DMFT framework, we find that the self-consistent equations bring in the lattice influence on the conduction electron bath, causing a pseudogap structure of the $t$-matrix that is lacking in the direct treatment of the KLM with independent bath approximation \cite{Wang2020_a}. Our calculations of the thermodynamic and transport properties also reproduce the desired properties qualitatively. Our method may be further extended to clusters \cite{Maier2005} to investigate the magnetic transition and quantum criticality for the KLM.

\section{Method}
We start with the Kondo lattice model,
\begin{equation}
H=\sum_{{\bf k}\alpha\nu}\epsilon_{{\bf k}}c^{\dagger}_{{\bf k}\alpha\nu}c_{{\bf k}\alpha\nu}+J_{K}\sum_{i}\mathbf{S}_{i}\cdot\mathbf{s}_{i},
\end{equation}
where $c^{\dagger}_{{\bf k}\alpha\nu}$ is the creation operator of a conduction electron with momentum $\bf k$, channel index $\nu\in[1,K]$ and spin index $\alpha\in[1,N]$, $\mathbf{s}_{i}$ is its spin operator, $\mathbf{S}_{i}$ is that of the local spins, and $J_K$ is their Kondo coupling. Our Schwinger boson approach enlarges the SU($2$) group to the SU($N$) group and represents the local spins as $S_{i,\alpha\alpha'}=b^{\dagger}_{i\alpha}b_{i\alpha'}-\delta_{\alpha\alpha'}\frac{2S}{N}$, in which $b^{\dagger}_{i\alpha}$ creates a bosonic spinon at site $i$ with spin index $\alpha$. A constraint is then imposed to ensure the physical subspace on the local sites, $n_{b}(i)\equiv \sum_{\alpha}b^{\dagger}_{i\alpha}b_{i\alpha}=2S$, which may be implemented by introducing a Lagrange multiplier on each site, $\sum_{i}\lambda_{i}(n_{b}(i)-2S)$. There are three distinct regimes depending on the ratio of $2S/K$: the underscreened ($2S/K>1$), overscreened ($2S/K<1$) and exactly-screened ($2S/K=1$) \cite{Parcollet1997}. In this work, we focus only on the exactly-screened case, $2S=K$, and fix the ratio $\kappa=2S/N$ that measures the quantum zero-point fluctuations \cite{Coleman2010} and may become important if nonlocal spatial correlations are considered \cite{Wang2020_b}.

Under the DMFT framework, the SU($N$) KLM is mapped to an effective Kondo impurity model with the action,
\begin{equation}
\begin{split}
S_{\rm eff}=&-\int^{\beta}_{0}d\tau d\tau'\sum_{\alpha\nu}\psi^{\dagger}_{\alpha\nu}(\tau)\mathcal{G}^{-1}_{0}(\tau-\tau')\psi_{\alpha\nu}(\tau')\\
&+\int^{\beta}_{0}d\tau \left[\sum_{\alpha}b^{\dagger}_{\alpha}(\tau)\left(\partial_{\tau}+\lambda\right)b_{\alpha}(\tau)-2S\lambda\right]\\
&+\frac{J_{K}}{N}\sum_{\alpha\alpha'\nu}\int^{\beta}_{0} d\tau\psi^{\dagger}_{\alpha'\nu}(\tau)\psi_{\alpha\nu}(\tau)b^{\dagger}_{\alpha}(\tau)b_{\alpha'}(\tau),
\end{split}
\end{equation}
where $\mathcal{G}^{-1}_{0}(\tau-\tau')$ plays the role of the effective Weiss field and $\psi^{\dagger}_{\alpha\nu}\equiv\frac{1}{\sqrt{\mathcal{N}_{s}}}\sum_{\bf k}c^{\dagger}_{{\bf k}\alpha\nu}$. $\mathcal{N}_{s}$ is the number of the lattice sites. The local Green's function and self-energy of conduction electrons satisfy the self-consistent equations:
\begin{align}
G^{\alpha\nu}_{c}(z)&=\int\frac{D(\epsilon)d\epsilon}{z+\mu-\epsilon-\Sigma^{\alpha\nu}_{c}(z)},\label{equation3}\\
\Sigma^{\alpha\nu}_{c}(z)&=\mathcal{G}_{0}^{-1}(z)-G^{\alpha\nu}_{c}(z)^{-1},
\end{align}
where $G^{\alpha\nu}_{c}(z)$ is the full local Green's function of conduction electrons on the lattice but also equal to that of the effective impurity model within the DMFT approximation, $\mu$ is the chemical potential, and $D(\epsilon)$ is the bare density of states (DOS). For square lattice, we use the dispersion $\epsilon_{\bf k}=-t(\cos k_{x}+\cos k_{y})$ with $t$ set to unity.

To apply the dynamical large-$N$ Schwinger boson approach as the impurity solver, we decouple the Kondo term using the Hubbard-Stratonovich transformation:
\begin{equation}
\begin{split}
&\frac{J_{K}}{N}\sum_{\nu\alpha\alpha'}\psi^{\dagger}_{\alpha'\nu}\psi_{\alpha\nu}b^{\dagger}_{\alpha}b_{\alpha'}\rightarrow\sum_{\nu\alpha}\bigg[\frac{1}{\sqrt{N}}(b^{\dagger}_{\alpha}\psi_{\alpha\nu})\chi_{\nu}\\
&+\frac{1}{\sqrt{N}}\chi^{\dagger}_{\nu}(\psi^{\dagger}_{\alpha\nu}b_{\alpha})\bigg]+\sum_{\nu}\frac{\chi^{\dagger}_{\nu}\chi_{\nu}}{J_{K}},
\end{split}
\end{equation}
where $\chi_{\nu}$ is an auxiliary Grassmann field representing a charged and spinless holon. The vertex $(b^{\dagger}_{\alpha}\psi_{\alpha\nu})\chi_{\nu}$ then yields an interaction between spinons, holons, and conduction electrons, causing their nontrivial dynamics. A set of self-energy equations can be derived using the generalized Luttinger-Ward functional $\Phi[G_{c},G_{\chi},G_{b}]$: $\Sigma^{\alpha\nu}_c(\tau)=\frac{\delta\Phi}{\delta G^{\alpha\nu}_c(-\tau)}$, $\Sigma^{\nu}_\chi(\tau)=\frac{\delta\Phi}{\delta G^{\nu}_\chi(-\tau)}$, and $\Sigma^{\alpha}_b(\tau)=-\frac{\delta\Phi}{\delta G^{\alpha}_b(-\tau)}$, where the plus (minus) sign before the differential indicates fermionic (bosonic) fields. In principle, $\Phi$ is the sum of all closed-loop two-particle irreducible skeleton Feynman diagrams \cite{Coleman2005_b}. But in the large-$N$ limit, only the leading-order diagram is retained, giving
\begin{equation}
\Phi[G_{c},G_{\chi},G_{b}]=\frac{1}{N}\sum_{\alpha\nu}\int^{\beta}_{0}G^{\alpha\nu}_{c}(\tau)G^{\alpha}_{b}(-\tau)G^{\nu}_{\chi}(\tau)d\tau.
\end{equation}
We have immediately
\begin{align}
\Sigma^{\alpha\nu}_{c}(\tau)&=\frac{1}{N}G^{\alpha}_{b}(\tau)G^{\nu}_{\chi}(-\tau),\label{equation7}\\
\Sigma^{\nu}_{\chi}(\tau)&=\frac{1}{N}\sum_{\alpha}G^{\alpha\nu}_{c}(-\tau)G^{\alpha}_{b}(\tau),\label{equation8}\\
\Sigma^{\alpha}_{b}(\tau)&=-\frac{1}{N}\sum_{\nu}G^{\alpha\nu}_{c}(\tau)G^{\nu}_{\chi}(\tau).\label{equation9}
\end{align}
The spinon and holon Green's functions (in frequency) are
\begin{align}
G^{\alpha}_{b}(z)&=\frac{1}{z-\lambda-\Sigma^{\alpha}_{b}(z)},\label{equation10}\\
G^{\nu}_{\chi}(z)&=\frac{1}{-1/J_{K}-\Sigma^{\nu}_{\chi}(z)},\label{equation11}
\end{align}
where $\lambda$ is a real constant in the saddle point approximation and will be tuned to satisfy the Schwinger boson constraint $\langle n_{b}\rangle=2S$.

The whole DMFT self-consistent procedures involve:
\begin{enumerate}
  \item [(1)] Setting initial self-energies $\Sigma^{\alpha\nu}_{c}$, $\Sigma^{\alpha}_{b}$, $\Sigma^{\nu}_{\chi}$ to zero,
  \item [(2)] Calculating $G^{\alpha\nu}_{c}$ via Eq. (\ref{equation3}),
  \item [(3)] Solving the effective impurity model:
  \begin{itemize}
  \item [(a)] Using Eq. (\ref{equation11}) to calculate $G^{\nu}_{\chi}$,
  \item [(b)] Using Eq. (\ref{equation10}) and the Schwinger boson constraint $\langle n_{b}\rangle=2S$ to calculate $\lambda$ and $G^{\alpha}_{b}$,
  \item [(c)] Using Eq. (\ref{equation8}-\ref{equation9}) to update $\Sigma^{\nu}_{\chi}$ and $\Sigma^{\alpha}_{b}$,
  \item [(d)] Repeating (a-c) until convergence,
  \end{itemize}
  \item [(4)] Using Eq. (\ref{equation7}) to update $\Sigma^{\alpha\nu}_{c}$,
  \item [(5)] Repeating (2-4) until convergence.
\end{enumerate}
The above equations can also be written in real frequency and solved self-consistently ($z\rightarrow\omega+i0^+$). In the following, we will drop all the spin/channel indices for simplicity due to the symmetry.

\begin{figure}[t]
\centering\includegraphics[width=0.51\textwidth]{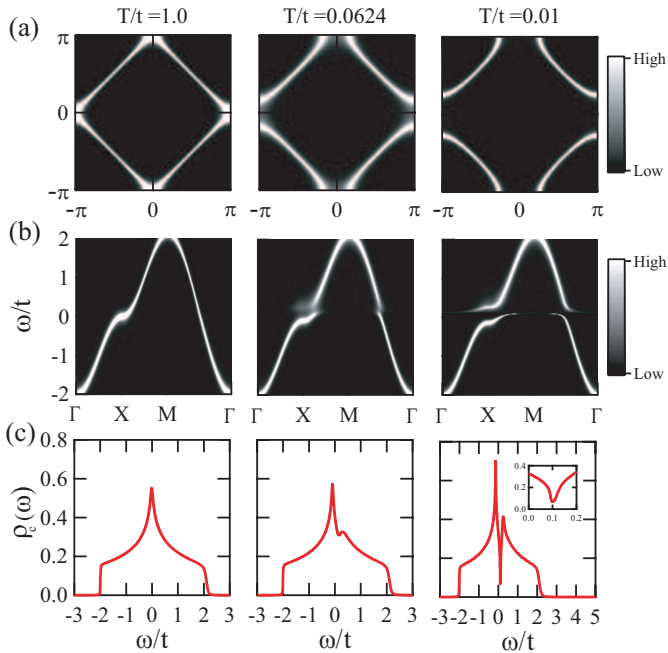}
\caption{Temperature evolution of (a) the Fermi surface, (b) the spectral function along the high symmetry path, and (c) the local density of states of conduction electrons. The inset highlights the pseudogap structure around the Fermi energy. The temperatures are chosen to be $T/t=1.0$, $0.0624$, $0.01$, corresponding to the high temperature limit, the Kondo temperature, and the Fermi liquid temperature, respectively. Other parameters are $J_{K}=1$, $\mu=0$, $K=2$ and $N=8$.}
\label{fig1}
\end{figure}

\section{Results and discussions}

Figure \ref{fig1}(a) shows the color plot of the conduction electrons' spectral function on a square lattice, $\rho_{c}({\bf k},\omega)=-{\rm Im}\left[1/(\omega-\epsilon_{\bf k}+\mu-\Sigma_{c}(\omega))\right]/\pi$, at the Fermi energy for three typical temperatures corresponding to the high temperature limit, the Kondo temperature and the Fermi liquid temperature, respectively. The white regions give the maximum of the spectral function and thus reflect the electron Fermi surface enclosing the area of negative $E_{\bf k}=\epsilon_{\bf k}+{\rm Re}\Sigma_{c}(0)$ \cite{Coleman2005_b}. Interestingly, although there is no explicit hybridization field and $f$ electron bands, we still see a gradual increase of the electron Fermi surface as the temperature decreases. Indeed, as shown in Fig. \ref{fig1}(b), a hybridization gap opens on the electron dispersion. It already emerges above the Kondo temperature but only becomes fully opened at lower temperatures, in agreement with the ARPES measurement on CeCoIn$_5$ \cite{Chen2017,Jang2020} and the suggested two-stage process by exact determinant quantum Monte Carlo simulations \cite{Hu2019} and the pump-probe experiment in CeCoIn$_5$ \cite{Liu2020PRL}. Correspondingly, a pseudogap structure is seen in Fig. \ref{fig1}(c) to develop in the local density of states $\rho_{c}(\omega)=-{\rm Im}G_{c}(\omega)/\pi$ slightly above the Fermi level, as also obtained by NRG \cite{Costi2002,Pruschke2000}. More results on the influence of other parameters are given in Appendix \ref{appendix1}.

\begin{figure}[t!]
\centering\includegraphics[width=0.5\textwidth]{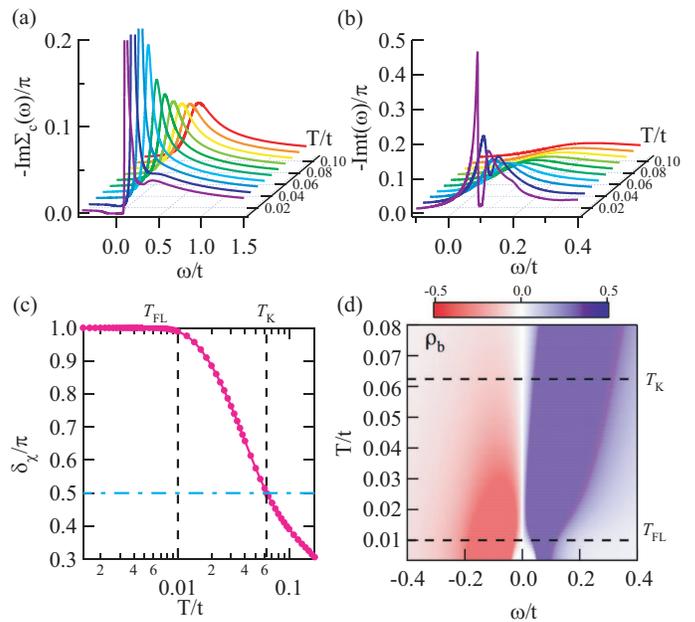}
\caption{ Temperature dependence of (a) the electron's self-energies, (b) the local $t$-matrix, (c) the holon phase shift, and (d) the spinon's spectral function. The Kondo temperature and the Fermi liquid temperature marked in (c) and (d) are determined from the holon phase shift. The parameters are the same as in Fig.~\ref{fig1}.}
\label{fig2}
\end{figure}

The hybridization physics is associated with a ``Kondo resonance peak" in the electron's self-energy $\Sigma_{c}$, as plotted in Fig. \ref{fig2}(a). Although $\Sigma_{c}$ is of the order of $1/N$ and thus disappears in the infinite-$N$ limit, it is essential for the appearance of the hybridization gap in the dispersion. At lower temperatures, a gap emerges in $\text{Im}\Sigma_{c}$ at the Fermi energy, signifying that the system enters the Fermi liquid state \cite{Rech2006}. The gap is, however, artificial. Within the Landau Fermi liquid theory, the imaginary part of $\Sigma_{c}$ is supposed to have a quadratic frequency dependence rather than a gap near the Fermi energy. How one can recover the quadratic frequency dependence is an open question in the Schwinger boson approach.

Our DMFT results are somewhat different from pure Schwinger boson calculations for the KLM. This is seen in the $t$-matrix of conduction electrons, which reflects the electron scattering off Kondo impurities and may be viewed as a correspondence of the $f$-electron spectral function in the Anderson lattice/impurity model \cite{Costi2002,Otsuki2009_a}. By definition, we have $G_{c}({\bf k},\omega)=G_{c0}({\bf k},\omega)+G_{c0}({\bf k},\omega)t({\bf k},\omega)G_{c0}({\bf k},\omega)$,
where $G_{c0}({\bf k},\omega)$ is the bare Green's function of conduction electrons on the lattice and $G_{c}({\bf k},\omega)=[G_{c0}({\bf k},\omega)^{-1}-\Sigma_{c}(\omega)]^{-1}$ is the ${\bf k}$-dependent full Green's function. The $t$-matrix is then associated with the local self-energy: $t({\bf k},\omega)\equiv\Sigma_{c}(\omega)+G_{c}({\bf k},\omega)\Sigma^{2}_{c}(\omega)$, so that the $\textbf{k}$-averaged local $t$ matrix is given by
\begin{equation}
t(\omega)\equiv\frac{1}{\mathcal{N}_{s}}\sum_{\bf k}t({\bf k},\omega)=\Sigma_{c}(\omega)+G_{c}(\omega)\Sigma^{2}_{c}(\omega). \label{t_matrix}
\end{equation}
Thus, within the Schwinger boson approach, the local $t$-matrix can only be obtained at finite-$N$ \cite{Lebanon2006,Lebanon2007}. Figure \ref{fig2}(b) plots its temperature dependence, where we find a double-peak structure that is missing in the pure Schwinger boson calculations for the KLM with independent electron baths \cite{Wang2020_a}. This structure is the fingerprint of the hybridization physics \cite{Grewe1991,Florens2004} and differs from the single resonance peak in the impurity model (Appendix \ref{appendix2}) \cite{Lebanon2006}. It arises from the second term of Eq. (\ref{t_matrix}) and thus contains information from the lattice through DMFT self-consistency. At high temperatures ($T\gg T_{\rm K}$), the hybridization is weak so that the first term dominates and gives a single peak in the $t$-matrix. Below $T_{\rm K}$, the lattice effect becomes pronounced and the second term gradually increases, causing the pseudogap above the Fermi level in Fig. \ref{fig2}(b).

It should be noted that the particle-hole symmetry is missing in the SU($N$) Schwinger boson formalism of the Kondo lattice model even for $\mu=0$ \cite{Saremi2007,Fritz2006,Vojta2001}. Rather, a pseudogap emerges above the Fermi level reflecting the hybridization physics at low temperatures. In the large-$N$ model with $N>2$ spin flavors, the particle-hole transformation of conduction electrons, $\psi_{i\alpha}\rightarrow \eta_{i}\psi_{i\alpha}^{\dagger}$ where $\eta_{i}=1(-1)$ on A(B) sublattice, requires a simultaneous transformation of local spins, $\mathbf{S}_i\rightarrow -\mathbf{S}_i$, to keep the Kondo Hamiltonian unchanged. For fermionic representation, this can be achieved by the particle-hole transformation of pseudofermions, but for the SU($N$) Schwinger boson representation, one cannot find such a transformation under the Schwinger boson constraint for $N>2$.

Two basic temperature scales, the Kondo temperature $T_{\rm K}$ and the Fermi liquid temperature $T_{\rm FL}$, can be identified from the holon phase shift calculated for the effective impurity model using
\begin{equation}
\delta_{\chi}=-{\rm Im}\ln\left[1/J^{*}_{K}+i\Sigma''_{\chi}(0) \right],
\end{equation}
where we have defined an effective Kondo coupling through $1/J^{*}_{K}\equiv 1/J_{K} + {\rm Re}\Sigma_{\chi}(0)$ \cite{Coleman2005_b}. At $T_{\rm K}$, $J^{*}_{K}$ changes sign and, correspondingly, the holon phase shift is equal to $\pi/2$. At $T_{\rm FL}$, $J^{*}_{K}<0$ and the imaginary part of $\Sigma_{\chi}$ is gapped, so the holon phase shift saturates to $\pi$. Figure \ref{fig2}(c) plots the holon phase shift as a function of temperature, where the two temperatures are indicated. The temperature-dependent evolution of spinon density of states $\rho_{b}(\omega)=-\frac{1}{\pi}{\rm Im}G_{b}(\omega)$ is also shown in Fig. \ref{fig2}(d) for comparison. We see also the formation of a gap below $T_{\rm FL}$, indicating the spinon confinement and full Kondo screening.

\begin{figure}[t!]
\centering\includegraphics[width=0.49\textwidth]{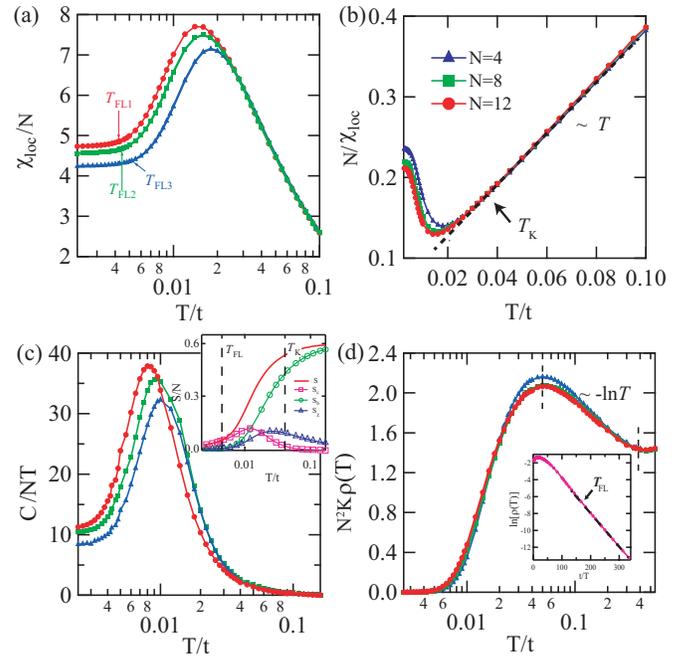}
\caption{Temperature evolution of (a) the static local magnetic susceptibility and (b) its inverse, (c) the specific heat coefficients $C/NT$, and (d) the resistivity $\rho(T)$, for $\kappa=0.25$ and $N=4$, $8$, $12$. The arrows indicate the Kondo temperature and the Fermi liquid temperature. The dashed line in (b) is a fit to the Curie-Weiss law. The inset of (c) shows the entropy and its three components for $N=8$. The inset of (d) gives the log-plot of the resistivity versus $t/T$ for $N=8$, showing artificial activation behavior (dashed line) due to spinon and holon gaps below the Fermi liquid temperature.}
\label{fig3}
\end{figure}

Within the DMFT and Schwinger boson framework, we may also calculate the temperature dependence of thermodynamic and transport properties. To avoid the influence of van Hove singularities, we assume in the following the hypercubic $D(\epsilon)=\exp(-\epsilon^{2}/2t^{2})/\sqrt{2\pi t^{2}}$ with $t=1$. The static local magnetic susceptibility $\chi_{\rm loc}(T)$ of the effective impurity model is given by
\begin{equation}
\frac{\chi_{\rm loc}(T)}{N}=\int \frac{d\omega}{\pi}n_{b}(\omega){\rm Im}\left[G_{b}^{2}(\omega)\right].
\end{equation}
The results are plotted in Fig. \ref{fig3}(a) and their inverse in Fig. \ref{fig3}(b). We see the typical Curie-Weiss behavior above $T_{\rm K}$ followed by a broad peak at lower temperatures. The deviation reflects the effect of the Kondo screening. Below $T_{\rm FL}$, $\chi_{\rm loc}(T)$ exhibits the typical Pauli susceptibility in the Fermi liquid state.

The entropy $S$ contains three components, $S/N=S_{c}+S_{b}+S_{\chi}$, with
\begin{equation}
\begin{split}
S_{c}&=-\frac{K}{\mathcal{N}_{s}}\sum_{\bf k}\int\frac{d\omega}{\pi}\frac{\partial n_{f}}{\partial T}\left[{\rm Im}\ln\left(\frac{G_{c0}}{G_{c}}\right)+G'_{c}\Sigma''_{c}\right],\\
S_{b}&=-\int\frac{d\omega}{\pi}\frac{\partial n_{b}}{\partial T}\left[{\rm Im}\ln(-G^{-1}_{b})+G'_{b}\Sigma''_{b}\right],\\
S_{\chi}&=-\kappa\int\frac{d\omega}{\pi}\frac{\partial n_{f}}{\partial T}\left[{\rm Im}\ln(-G_{\chi}^{-1})+G'_{\chi}\Sigma''_{\chi}\right],
\end{split}
\end{equation}
where $n_{f/b}$ is the Fermi/Bose distribution function. The specific heat coefficient is given by $C/T=dS/dT$ and plotted in Fig. \ref{fig3}(c). Above $T_{\rm K}$, the local moments are free and the entropy (inset) approaches its high temperature value $S_{{\rm high\text{-}T}}/N=(1+\kappa)\ln(1+\kappa)-\kappa\ln(\kappa)=0.625$. Below $T_{\rm K}$, the local moments are gradually screened; $C/NT$ increases rapidly with lowering temperature and then exhibits a broad ``Schottky" peak. In the Fermi liquid state, spinons and holons are both confined and we find a constant specific heat coefficient as is expected for a Landau Fermi liquid.

The resistivity $\rho(T)$ may also be calculated using the linear response theory,
\begin{equation}
\rho(T)=\frac{1}{NK}\left[\pi\int d\epsilon d\omega D(\epsilon) \rho^{2}_{c}(\epsilon,\omega)\left(-\frac{dn_{f}}{d\omega}\right) \right]^{-1}.
\end{equation}
A logarithmic temperature dependence is also seen for the resistivity in Fig. \ref{fig3}(d), reflecting incoherence Kondo scattering above $T_{\rm K}$. Below $T_{\rm K}$, the resistivity starts to exhibit metallic behavior. The broad maximum thus corresponds to the coherence peak as seen in typical heavy fermion metals \cite{Jang2020}. It should be noted that the Schwinger boson approach cannot produce the correct Fermi-liquid behavior with $T^2$ dependence. As shown in the inset of Fig. \ref{fig3}(d), an artificial activation behavior appears due to finite spinon and holon gaps below the Fermi liquid temperature. Hence, higher-order diagrams of $\Sigma_{c}$ are needed in order to recover the correct $T^{2}$ scaling in the resistivity \cite{Lebanon2006,Wang2020_a}. Other than that, our results are in qualitative agreement with the expectation from more rigorous calculations, establishing the validity of the Schwinger boson approach as a potentially useful impurity solver for DMFT.

\section{Conclusion}
To summarize, we have applied the large-$N$ Schwinger boson approach as an impurity solver for the DMFT calculations of the KLM. Our method can work on the real-frequency axis and give qualitatively correct results over a wide range of temperature. It yields the hybridization physics from non-vanishing conduction electron self-energies and the pesudogap structure in the $t$-matrix from DMFT self-consistency. The resulting thermodynamic and transport properties provide a good description of the crossover from the local moment to Fermi liquid regimes. Take together, we expect that it may be further extended to cluster DMFT or multi-orbital models and eventually combine with the density-functional theory as an alternative efficient method for real material calculations.\\

\acknowledgements
This work was supported by the National Natural Science Foundation of China (NSFC Grant No. 11974397,  No. 11774401, No. 12174429), the National Key R\&D Program of MOST of China (Grant No. 2017YFA0303103), the Strategic Priority Research Program of the Chinese Academy of Sciences (Grant No. XDB33010100), and the Youth Innovation Promotion Association of CAS.\\

\appendix

\section{Influence of other parameters} \label{appendix1}
We provide here more data on the influence of the values of $N$, $\mu$, and $2S$.

\begin{figure}[h]
\centering\includegraphics[width=0.49\textwidth]{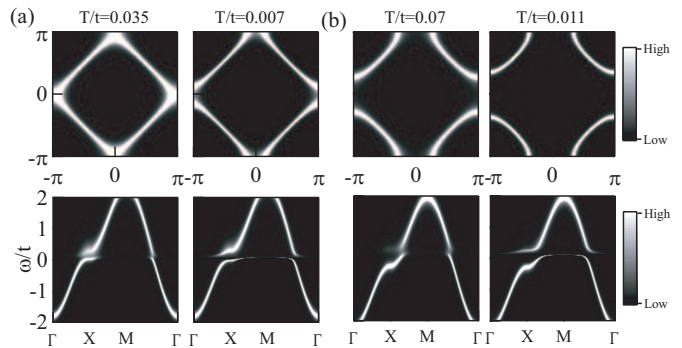}
\caption{The Fermi surface and spectral function along the high symmetry path for (a) $\mu=-0.2$ and (b) $\mu=0.2$. The temperatures are $T/t=0.035$, $0.007$ for (a) and $T/t=0.07$, $0.011$ for (b), corresponding to their respective Kondo and the Fermi liquid temperatures. Other parameters are $J_K=1$, $K= 2$, and $N=8$.}
\label{fig4}
\end{figure}

Figure \ref{fig4} compares the results for different chemical potentials, $\mu=\pm 0.2$, at their respective Kondo and Fermi liquid temperatures. We find both $T_{\rm K}$ and $T_{\rm FL}$ decrease with lowering $\mu$ or $n_c$, in agreement with previous calculations based on the slave boson approach \cite{Burdin2000} and NRG \cite{Pruschke2000}.

\begin{figure}[b]
\centering\includegraphics[width=0.49\textwidth]{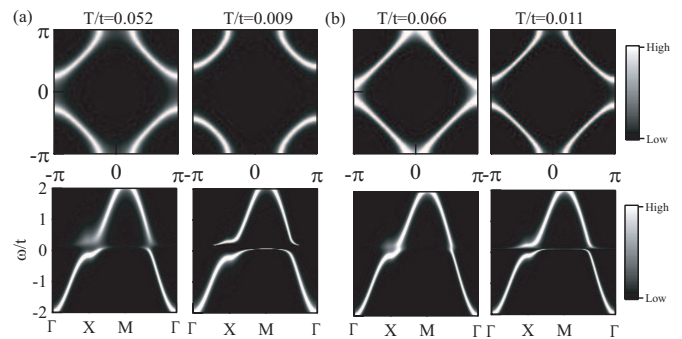}
\caption{The Fermi surface and spectral function along the high symmetry path for (a) $N=4$ and $T/t=0.052$, $0.009$; (b) $N=12$ and $T/t=0.066$, $0.011$. The temperatures are chosen according to their respective Kondo and Fermi liquid temperatures. Other parameters are $\kappa=0.25$, $J_K=1$, and $\mu=0$.}
\label{fig5}
\end{figure}

Figure \ref{fig5} compares the Fermi surface and spectral function for $N=4$ and 12 at their respective Kondo and Fermi liquid temperatures. Both the hybridization gap and the Fermi surface expansion at low temperatures are qualitatively unchanged. However, the volume of the Fermi surface $v_{Fs}$ shrinks as $N$ increases because of the Luttinger sum rule \cite{Coleman2005_b}:
\begin{equation}
\frac{v_{Fs}}{(2\pi)^{d}}=\frac{\sum_{{\bf k}\alpha\nu}c^{\dagger}_{{\bf k}\alpha\nu}c_{{\bf k}\alpha\nu}}{NK}+\frac{1}{N}=n_{c}+\frac{1}{N},
\end{equation}
where $n_{c}$ is the density of conduction electrons. In the Fermi liquid state, the volume is expanded by the size of $1/N$ for each of the $NK$-fold degenerate bands to incorporate a  local spin of the size $2S=K$.

\begin{figure}[h!]
\centering\includegraphics[width=0.49\textwidth]{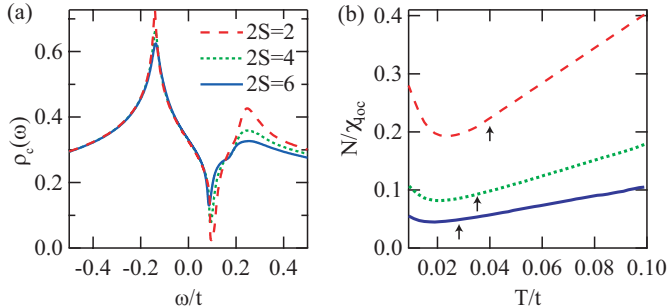}
\caption{Comparison of (a) the local density of states of conduction electrons and (b) the inverse local magnetic susceptibility for $2S=2$, 4, 6. The arrows in (b) mark the respective Kondo temperature. Other parameters are $J_K=1$, $\mu=0$, and $N=8$.}
\label{fig6}
\end{figure}

Figure \ref{fig6}(a) shows the influence of $2S$ on the conduction electrons' local density of states. The pseudogap always exists but its location, which is primarily associated with the Kondo temperature $T_{\rm K}$ \cite{Wang2020_a}, moves slightly towards the Fermi energy with increasing $2S$. This reflects the reduction of $T_{\rm K}$ for large spin size, where the local spin behaves more like a classical spin, as also marked in Fig. \ref{fig6}(b) by the deviation of the inverse local magnetic susceptibility from the high-temperature Curie-Weiss law.

\section{Comparison of Kondo lattice and Kondo impurity models} \label{appendix2}
\begin{figure}[h!]
\centering\includegraphics[width=0.49\textwidth]{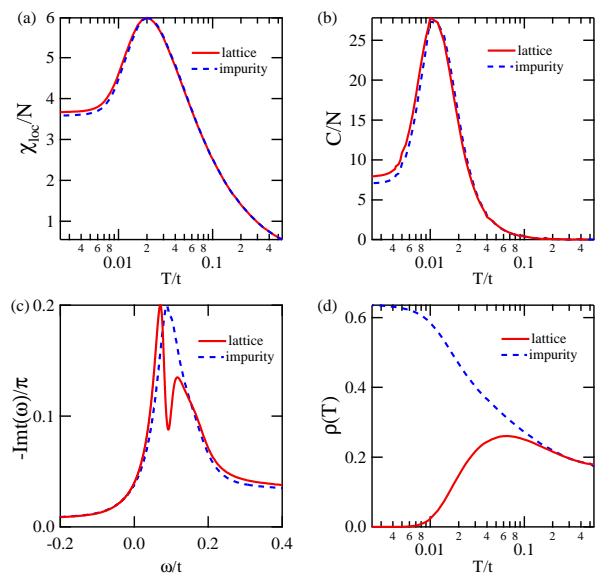}
\caption{Comparison of (a) the local magnetic susceptibility, (b) the specific heat, (c) the local $t$-matrix, and (d) the resistivity for the Kondo lattice model (solid line) and the Kondo impurity model (dashed line). The parameters are $J_K=1$, $K= 2$, and $N=8$ for both models. }
\label{fig7}
\end{figure}

Here we provide some detailed comparison between the Kondo impurity model and our results on the Kondo lattice model. Figures \ref{fig7}(a) and \ref{fig7}(b) compare their local susceptibility and specific heat, where conduction electron contributions have been subtracted. No significant differences are seen since both are governed by local properties. This may explain why the single-impurity model has often been used in experimental literatures for fitting the data on a Kondo lattice compound.

On the other hand, qualitative distinctions can be clearly seen in Fig. \ref{fig7}(c) for the local $t$-matrix and Fig. \ref{fig7}(d) for the resistivity. The $t$-matrix may be tentatively associated with the $f$-electron spectral function in the Anderson model. We see that it develops a single peak for the Kondo impurity model but a double-peak structure for the Kondo lattice model. The former manifests the Kondo resonance in the impurity model, while the latter reflects the hybridization physics in the lattice. At low temperatures, the resistivity saturates for the impurity model but decreases to nearly zero for the lattice, implying a fundamental distinction between two models, namely, unitary scattering for fully screened Kondo impurity and the development of a coherent heavy electron state in the Kondo lattice.

\begin{thebibliography}{20}
\bibitem{Hewson1997}A. C. Hewson, \emph{The Kondo Problem to Heavy Fermions}, (Cambridge University Press, Cambridge, England, 1997).
\bibitem{Coleman2005_a}P. Coleman, \emph{Introduction to Many-Body Physics}, (Cambridge University Press, Cambridge, England, 2015).
\bibitem{Metzner1989} W. Metzner and D. Vollhardt, Correlated Lattice Fermions in $d=\infty$ Dimensions, Phys. Rev. Lett. {\bf 62}, 324 (1989).
\bibitem{Georges1992} A. Georges, G. Kotliar, and Q. Si, Strongly correlated systems in infinite Dimensions and their zero Dimensional counterparts, Int. J. Mod. Phys. B {\bf 6}, 705 (1992).
\bibitem{Georges1996} A. Georges, G. Kotliar, W. Krauth, and M. J. Rozenberg, Dynamical mean-field theory of strongly correlated fermion systems, Rev. Mod. Phys. {\bf 68}, 13 (1996).
\bibitem{Costi2002} T. A. Costi and N. Manini, Low Energy Scales and Temperature-Dependent Photoemission of Heavy Fermions, J. Low Temp. Phys. {\bf 126}, 835 (2002).
\bibitem{Otsuki2009_a} J. Otsuki, H. Kusunose, and Y. Kuramoto, The Kondo Lattice Model in Infinite Dimensions: \uppercase\expandafter{\romannumeral1}. Formalism, J. Phys. Soc. Jpn. {\bf 78}, 014702 (2009).
\bibitem{Otsuki2009_b} J. Otsuki, H. Kusunose, and Y. Kuramoto, The Kondo Lattice Model in Infinite Dimensions: \uppercase\expandafter{\romannumeral2}. Static Susceptibilities and Phase Diagram, J. Phys. Soc. Jpn. {\bf 78}, 034719 (2009).
\bibitem{Bodensiek2011} O. Bodensiek, R. \v{Z}itko, R. Peters, and T. Pruschke, Low-energy properties of the Kondo lattice model, J. Phys.: Condens. Matter {\bf 23}, 094212 (2011).
\bibitem{Matsumoto1995} N. Matsumoto and F. J. Ohkawa, Kondo-lattice model in infinite dimensions, Phys. Rev. B {\bf 51}, 4110 (1995).
\bibitem{Gull2012} E. Gull, A. J. Millis, A. I. Lichtenstein, A. N. Rubtsov, M. Troyer, and P. Werner, Continuous-time Monte Carlo methods for quantum impurity models, Rev. Mod. Phys. {\bf 83}, 349 (2011).
\bibitem{Wilson1975} K. G. Wilson, The renormalization group: Critical phenomena and the Kondo problem, Rev. Mod. Phys. {\bf 47}, 773 (1975).
\bibitem{Bulla2008} R. Bulla, T. A. Costi, and T. Pruschke, Numerical renormalization group method for quantum impurity systems, Rev. Mod. Phys. {\bf 80}, 395 (2008).
\bibitem{Kim1990} C. I. Kim, Y. Kuramoto, and T. Kasuya, Self-consistent dynamical theory for the Anderson lattice, J. Phys. Soc. Jpn. {\bf 59}, 2414 (1990).
\bibitem{Palsson2001} G. P\'{a}lsson, Computational studies of thermoelectricity in strongly correlated electron systems, Ph.D. thesis, Rutgers University, 2001.
\bibitem{Arovas1988} D. P. Arovas and A. Auerbach, Functional integral theories of low-dimensional quantum Heisenberg models, Phys. Rev. B {\bf 38}, 316 (1988).
\bibitem{Parcollet1997} O. Parcollet and A. Georges, Transition from Overscreening to Underscreening in the Multichannel Kondo Model: Exact Solution at Large $N$, Phys. Rev. Lett. {\bf 79}, 4665 (1997).
\bibitem{Rech2006} J. Rech, P. Coleman, G. Zarand, and O. Parcollet, Schwinger Boson Approach to the Fully Screened Kondo Model, Phys. Rev. Lett. {\bf 96}, 016601 (2006).
\bibitem{Komijani2018} Y. Komijani and P. Coleman, Model for a Ferromagnetic Quantum Critical Point in a 1D Kondo Lattice, Phys. Rev. Lett. {\bf 120}, 157206 (2018).
\bibitem{Komijani2019} Y. Komijani and P. Coleman, Emergent Critical Charge Fluctuations at the Kondo Breakdown of Heavy Fermions, Phys. Rev. Lett. {\bf 122}, 217001 (2019).
\bibitem{Wang2020_a} J. Wang, Y.-Y. Chang, C.-Y. Mou, S. Kirchner, and C.-H. Chung, Quantum phase transition in a two-dimensional Kondo-Heisenberg model: A dynamical Schwinger-boson large-$N$ approach, Phys. Rev. B {\bf 102}, 115133 (2020).
\bibitem{Wang2020_b} J. Wang and Y.-F. Yang, Nonlocal Kondo effect and quantum critical phase in heavy-fermion metals, Phys. Rev. B {\bf 104}, 165120 (2021).
\bibitem{Drouin-Touchette2021} V. Drouin-Touchette, E. J. K\"{o}nig, Y. Komijani, and P. Coleman, Emergent moments in a Hund's impurity, Phys. Rev. B {\bf 103}, 205147 (2021).
\bibitem{Shen2020}  B. Shen, Y. Zhang, Y. Komijani, M. Nicklas, R. Borth, A. Wang, Y. Chen, Z. Nie, R. Li, X. Lu, H. Lee, M. Smidman, F. Steglich, P. Coleman, and H. Q. Yuan, Strange-metal behaviour in a pure ferromagnetic Kondo lattice, Nature(London) {\bf 579}, 51 (2020).
\bibitem{Zhao2019} H. Zhao, J. Zhang, M. Lyu, S. Bachus, Y. Tokiwa, P. Gegenwart, S. Zhang, J. Cheng, Y.-F. Yang, G. Chen, Y. Isikawa, Q. Si, F. Steglich, and P. Sun, Quantum-critical phase from frustrated magnetism in a strongly correlated metal, Nat. Phys. {\bf 15}, 1261 (2019).
\bibitem{Kotliar2006} G. Kotliar, S. Y.  Savrasov, K. Haule, V. S. Oudovenko, O. Parcollet, and C. A. Marianetti, Electronic structure calculations with dynamical mean-field theory, Rev. Mod. Phys. {\bf 78}, 865 (2006).
\bibitem{Auerbach1986} A. Auerbach and K. Levin, Kondo Bosons and the Kondo Lattice: Microscopic Basis for the Heavy Fermi Liquid, Phys. Rev. Lett. {\bf 57}, 877 (1986).
\bibitem{Read1983} N. Read and D. M. Newns, On the solution of the Coqblin-Schreiffer Hamiltonian by the large-$N$ expansion technique, J. Phys. C {\bf 16}, 3273 (1983).
\bibitem{Maier2005} T. Maier, M. Jarrell, T. Pruschke, and M. H. Hettler, Quantum cluster theories, Rev. Mod. Phys. {\bf 77}, 1027 (2005).
\bibitem{Coleman2010} P. Coleman and A. H. Nevidomskyy, Frustration and the Kondo Effect in Heavy Fermion Materials, J. Low. Temp. Phys. {\bf 161}, 182 (2010).
\bibitem{Coleman2005_b} P. Coleman, I. Paul, and J. Rech, Sum rules and Ward identities in the Kondo lattice, Phys. Rev. B {\bf 72}, 094430 (2005).
\bibitem{Chen2017} Q. Y. Chen, D. F. Xu, X. H. Niu, J. Jiang, R. Peng, H. C. Xu, C. H. P. Wen, Z. F. Ding, K. Huang, L. Shu, Y. J. Zhang, H. Lee, V. N. Strocov, M. Shi, F. Bisti, T. Schmitt, Y. B. Huang, P. Dudin, X. C. Lai, S. Kirchner, H. Q. Yuan, and D. L. Feng, Direct observation of how the heavy-fermion state develops in CeCoIn$_5$, Phys. Rev. B {\bf 96}, 045107 (2017).
\bibitem{Jang2020}  S. Y. Jang, J. D. Denlinger, J. W. Allen, V. S. Zapf, M. B. Maple, J. N. Kim, B. G. Jang, and J. H. Shim, Evolution of the Kondo lattice electronic structure above the transport coherence temperature, Proc. Natl. Acad. Sci. U.S.A. {\bf 117}, 23467 (2020).
\bibitem{Hu2019} D. Hu, J.-J Dong, and Y.-F. Yang, Hybridization fluctuations in the half-filled periodic Anderson model, Phys. Rev. B {\bf 100}, 195133 (2019).
\bibitem {Liu2020PRL} Y. P. Liu, Y. J. Zheng, J.-J Dong, H. Lee, Z. X. Wei, W. L. Zhang, C. Y. Chen, H. Q. Yuan, Y.-F. Yang, and J. Qi, Hybridization Dynamics in CeCoIn$_5$ Revealed by Ultrafast Optical Spectroscopy, Phys. Rev. Lett. \textbf{124}, 057404 (2020).
\bibitem{Pruschke2000} T. Pruschke, R. Bulla, and M. Jarrell, Low-energy scale of the periodic Anderson model, Phys. Rev. B {\bf 61}, 12799 (2000).
\bibitem{Lebanon2006} E. Lebanon, J. Rech, P. Coleman, and O. Parcollet, Conserving Many Body Approach to the Infinite-$U$ Anderson Model, Phys. Rev. Lett. {\bf 97}, 106604 (2006).
\bibitem{Lebanon2007} E. Lebanon and P. Coleman, Fermi liquid identities for the infinite-$U$ multichannel Anderson model, Phys. Rev. B {\bf 76}, 085117 (2007).
\bibitem{Grewe1991} N. Grewe and F. Steglich, in \emph{Handbook on the Physics and Chemistry of Rare Earths}, edited by K. A. Gschneidner and L. Eyring (North-Holland, Amsterdam, 1991).
\bibitem{Florens2004} S. Florens, Singular dynamics and pseudogap formation in the underscreened Kondo impurity and Kondo lattice models, Phys. Rev. B {\bf 70}, 165112 (2004).
S. Saremi and P. A. Lee, Quantum critical point in the Kondo-Heisenberg model on the honeycomb lattice, Phys. Rev. B {\bf 75}, 165110 (2007).
\bibitem{Fritz2006} L. Fritz, S. Florens, and M. Vojta, Universal crossovers and critical dynamics of quantum phase transitions: A renormalization group study of the pseudogap Kondo problem, Phys. Rev. B {\bf 74}, 144410 (2006).
\bibitem{Vojta2001} M. Vojta, Multichannel Pseudogap Kondo Model: Large-$N$ Solution and Quantum-Critical Dynamics, Phys. Rev. Lett. {\bf 87}, 097202 (2001).
\bibitem{Burdin2000}S. Burdin, A. Georges, and D. R. Grempel, Coherence Scale of the Kondo Lattice, Phys. Rev. Lett. {\bf 85}, 1048 (2000).
\end{thebibliography}
\end{document}